\documentclass[preprintnumbers, floatfix, preprintnumbers, letterpaper, superscriptaddress,nofootinbib,twocolumn]{revtex4}
\pdfoutput=1
\usepackage{graphicx}
\usepackage{microtype}
\usepackage{amsmath}
\usepackage{amssymb}
\usepackage{subfigure}
\usepackage{hyperref}
\usepackage{url}
\usepackage{xcolor}
\usepackage{color}
\usepackage{mathrsfs}
\usepackage{calrsfs}
\usepackage{amsfonts}
\usepackage{latexsym}
\usepackage{ragged2e}
\usepackage{epsfig}
\usepackage{textcomp}

\usepackage{caption}
\DeclareCaptionJustification{justified}{\leftskip=0pt \rightskip=0pt \parfillskip=0pt plus 1fil}
\definecolor{vividviolet}{rgb}{0.62, 0.0, 1.0}
\definecolor{amaranth}{rgb}{0.9, 0.17, 0.31}
\definecolor{palatinateblue}{rgb}{0.15, 0.23, 0.89}
\definecolor{brightpink}{rgb}{1.0, 0.0, 0.5}
\definecolor{cornflowerblue}{rgb}{0.39, 0.58, 0.93}
\definecolor{deepcarminepink}{rgb}{0.94, 0.19, 0.22}
\definecolor{radicalred}{rgb}{1.0, 0.21, 0.37}

\hypersetup{ linktoc=all,
    colorlinks, linkcolor={palatinateblue},
    citecolor={brightpink}, urlcolor={amaranth}
}

\graphicspath{{Images/}}

\renewcommand{\d}[1]{\ensuremath{\operatorname{d}\!{#1}}}

\def\sideremark#1{\ifvmode\leavevmode\fi\vadjust{\vbox to0pt{\vss% the remark
 \hbox to 0pt{\hskip\hsize\hskip1em%                          will appear only
 \vbox{\hsize1.2cm\tiny\raggedright\pretolerance10000%          on the side
 \noindent #1\hfill}\hss}\vbox to8pt{\vfil}\vss}}}%
\begin{document}

\title{No Violation of the Second Law in Extended Black Hole Thermodynamics\footnote{
Alberta-Thy-5-19}
}
\author{Shi-Qian \surname{Hu}}
\email{mx120170256@yzu.edu.cn}
\affiliation{Center for Gravitation and Cosmology, College of Physical Science and Technology, Yangzhou University, \\180 Siwangting Road, Yangzhou City, Jiangsu Province  225002, China}

\author{Yen Chin \surname{Ong}}
\email{ycong@yzu.edu.cn}
\affiliation{Center for Gravitation and Cosmology, College of Physical Science and Technology, Yangzhou University, \\180 Siwangting Road, Yangzhou City, Jiangsu Province  225002, China}
\affiliation{School of Aeronautics and Astronautics, Shanghai Jiao Tong University, Shanghai 200240, China}
%\affiliation{Nordita, KTH Royal Institute of Technology \& Stockholm University,
%Roslagstullsbacken 23, SE-106 91 Stockholm, Sweden}

\author{Don N. \surname{Page}}
\email{profdonpage@gmail.com}
\affiliation{Department of Physics,Theoretical Physics Institute, 4-181 CCIS,\\ University of Alberta, Edmonton, Alberta T6G 2E1, Canada}

\begin{abstract} Recently a number of papers have claimed that the horizon area -- and thus the entropy -- of near extremal black holes in anti-de Sitter spacetimes can be reduced by dropping particles into them. In this note we point out that this is a consequence of an underlying assumption that the energy of an infalling particle changes only the “internal energy” of the black hole, whereas a more physical assumption would be that it changes the enthalpy (mass). In fact, under the latter choice, the second law of extended black hole thermodynamics is no longer violated.
\end{abstract}

\maketitle
\section{Introduction: Extended Black Hole Thermodynamics and the Second Law}\label{1}

In 1973 \cite{bch}, Bardeen, Carter, and Hawking discovered that in asymptotically flat Kerr-Newman black hole spacetimes in general relativity, the black hole parameters satisfy equations that appear to be analogous to the laws of thermodynamics. Subsequently, with the discoveries of the Bekenstein-Hawking entropy and Hawking radiation, this so-called ``black hole mechanics'', the term coined in \cite{bch}, was gradually replaced by ``black hole thermodynamics''. That is to say, the viewpoint in the field has shifted from treating these properties of black holes as mere analogous curiosities to a \emph{bona fide} thermodynamics\footnote{A nice discussion on why black hole thermodynamics is indeed as real as ordinary thermodynamics, but only after Hawking radiation is incorporated, can be found in \cite{1710.02724, 1710.02725}.}.  This understanding has changed our understanding of black holes ever since, opening up vast areas of research including phase transitions and holography, but at the same time, also eventually led to the infamous information paradox. For a review of black hole thermodynamics, see e.g., \cite{1804.10610}.

For an asymptotically flat Kerr-Newman black hole, the usual first law of black hole mechanics takes the form
\begin{equation}
\d M= T\d S + \Phi\d Q + \Omega\d J,
\end{equation}
where $M$ denotes the ADM mass of the black hole, $T$ its Hawking temperature, $S$ its Bekenstein-Hawking entropy, $\Phi$ its electrostatic potential, $Q$ its electrical charge, $\Omega$ its angular velocity and $J$ its angular momentum. Note that there is an absence of a $V\d P$ term usually found in ordinary thermodynamics. This term in the context of black hole spacetime was eventually introduced \cite{1209.1272, 0904.2765}, which requires an anti-de Sitter (AdS) background, since the pressure $P$ is related to the cosmological constant $\Lambda$. More precisely, $P=-\Lambda/(8\pi G)$, where $G$ is Newton's gravitational constant and $\Lambda <0$. The ``thermodynamical volume'' $V$ is then defined as the thermodynamic conjugate $\partial M/\partial P$.  In this ``extended thermodynamics'', which was later dubbed ``black hole chemistry'' \cite{1404.2126v1}, the first law is generalized to
\begin{equation}
\d M= T\d S + \Phi\d Q + \Omega\d J + V\d P,
\end{equation}
where $M$ is now \emph{re-interpreted} as the \emph{enthalpy}, sometimes denoted by $H$, of the system:
\begin{equation}
M \equiv H:= U + PV.
\end{equation}
The $PV$ term in this equation can be interpreted as the contribution to the total mass-energy $M$ of the black hole from the work needed to exclude the volume $V$. 

We emphasize that $V$ has nothing to do with any geometrical notion of volume \emph{a priori}, though it turns out that for static charged or neutral black holes, 
\begin{equation}
V=\frac{4}{3}\pi r_+^3, 
\end{equation}
where $r_+$ denotes the outer horizon of the black hole, as if the black hole were a ball of radius $r_+$ in $\Bbb{R}^3$. This is no longer true for more complicated spacetimes such as AdS-Kerr or AdS-Taub-NUT; for the latter the thermodynamical volume is even negative \cite{1405.5941}!

Given that black hole chemistry seems to give plenty of physically reasonable results (for a review, see \cite{1608.06147}), it therefore came as a surprise when it was claimed by Gwak that by dropping a charged particle into a near-extremal black hole one could \emph{reduce} the area of the black hole horizon \cite{1709.08665}. Since the horizon area is interpreted as the entropy in black hole thermodynamics, this would mean that the second law has been violated. This result has received much attention lately and has been calculated to occur in a wide variety of black holes, going beyond general relativity and/or with various matter sources \cite{1901.04247, 1901.05589, 1901.08915, 1901.10660, 1902.06489, 1903.03764, 1904.12365,1905.01618,1905.07747,1905.07750}. Notably, the case without electrical charge but with rotation, i.e. AdS-Kerr, was investigated \cite{1905.01618}. The authors found that the validity of the second law depends on the spin parameter and the value of the cosmological constant. 

One might be tempted to treat this as a very special property peculiar to black holes in anti-de Sitter spacetimes, and argue that violations of the second law simply mean that these objects are not physically relevant. However, the utility of asymptotically AdS black holes is their wide applications in holography (now commonly also known as gauge/gravity duality, since the applicability of such correspondence has been demonstrated to go beyond the original ``AdS/CFT correspondence''). Thus any violation of the second law in the context of AdS black holes also translates to a violation of the second law in the corresponding ordinary quantum field theory, which is unacceptable\footnote{In the extended thermodynamics, allowing the pressure, or equivalently the cosmological constant, to change is equivalent to allowing the number of colors, $N$, in the corresponding boundary field theory to vary \cite{1404.5982, 1406.7267,1409.3521}.}. Of course, in a typical holography with finite temperature field theory, the Hawking temperature is not zero, and it is the generalized second law -- the total entropy of black hole and radiation, as with as any matter in the bulk -- that has to be non-decreasing. Nevertheless, it is alarming to see the violation of the second law at the classical level when Hawking radiation is not considered. Indeed we could consider a black hole whose outgoing Hawking radiation is balanced by ingoing radiation which was reflected back from the AdS boundary: then the entropy content of the radiation is approximately constant, so any change to the entropy of the system corresponds to the change to the black hole entropy. 

Given that there is now strong theoretical evidence in support on the notion of black hole chemistry, this putative violation of the second law \emph{must} be explained.
One possibility is that the notion of entropy must be modified, i.e. it is no longer given by the horizon of the area alone, but instead a hitherto unknown correction term exists that once it is properly included the second law can be restored. This point of view was already raised in \cite{1709.08665}. Another possibility is that the violation can somehow be prevented when backreaction is properly taken into account, much like the way apparent violations of cosmic censorship in the attempts to over-spin a black hole \cite{0907.4146} can often be prevented by considering the ``self-force'' of the particle \cite{0205005, 1008.5159, 1501.07330, Wald}.

However, our opinion is that this violation of the second law can be explained in a simpler manner: when a charged particle is dropped into a black hole, what changes is its enthalpy or mass\footnote{There is also a change in the electrical charge, as we will see in Sec. (\ref{2}), but this is not the main point to emphasize.}, \emph{not} just the internal energy as assumed in \cite{1709.08665} and the follow-up works. In Sec. (\ref{2}) we will review Gwak's argument in \cite{1709.08665} and explain how this changes the arguments and saves the second law. Finally in 
Sec. (\ref{3}) we will conclude with some discussions to explain the reasons why when a particle is dropped into a black hole, it is the enthalpy that should increase by the particle energy instead of the internal energy.

\section{The Change of Enthalpy Ensures Validity of the Second Law}\label{2}

Consider a Reissner-Nordstr\"om-AdS black hole, the original case studied in \cite{1709.08665}, whose metric tensor is given by (using units such that $G=c=4\pi \epsilon_0=1$)
\begin{equation}
\d s^2 = -f(r) \d t^2+ f(r)^{-1}\d r^2+ r^2\left(\d\theta^2 + \sin^2\theta \d\phi^2\right),
\end{equation}
in which (with $\Lambda=-3/L^2$)
\begin{equation}
f(r)=1-\frac{2M}{r}+\frac{Q^2}{r^2}+\frac{r^2}{L^2}.
\end{equation}
Consider a charged particle dropped into the black hole: then at the horizon, $r=r_+$, where $f(r_+)=0$, 
\begin{equation}\label{df}
\d f = \frac{\partial f}{\partial M}\d M + \frac{\partial f}{\partial Q}\d Q + \frac{\partial f}{\partial L} \d L + \frac{\partial f}{\partial r_+} \d r_+ =0.
\end{equation}
For simplicity, we will only focus on the 4-dimensional case, though the following naturally also generalizes to higher dimensions. It can be shown that Eq. (\ref{df}) becomes
\begin{equation}\label{df2}
\d M - \frac{Q}{r_+}\d Q + \frac{r_+^3}{L^3}\d L = \left(\frac{r_+^2}{L^2}+\frac{M}{r_+}-\frac{Q^2}{r_+^2}\right)\d r_+.
\end{equation}

In \cite{1709.08665}, Gwak considered dropping a charged particle with energy $E$ and charge $q$ into the black hole.
He showed that a standard argument via the Hamilton-Jacobi equation yields the relationship 
\begin{equation}
E=\frac{Q}{r_+}q+|p^r|,
\end{equation}
where $|p^r|$ denotes the radial momentum of the particle.
He then assumed that the energy of the particle changes the \emph{internal energy} of the black hole, i.e.
$E=\d U$, and hence
\begin{equation}\label{gwak}
\d M - \d(PV) = \frac{Q}{r_+}\d Q + |p^r|.
\end{equation}
Expanding $\d(PV)=V\d P + P \d V$, and substituting in the expressions for $V$ and $P$ in terms of the black hole parameters, one would obtain
\begin{equation}
\d M + \frac{r_+^3}{L^3}\d L - \frac{3}{2}\left(\frac{r_+^2}{L^2}\right)\d r_+ =\frac{Q}{r_+}\d Q + |p^r|.
\end{equation}
Substituting this into Eq. (\ref{df2}) and simplifying, one would find that the terms involving $\d Q$ and $\d L$ cancel out, and 
eventually, 
%\begin{equation}
%\d r_+ = \frac{2 r_+^3 L^2 |p^r|}{2Mr_+^2 L^2 - r_+^5 -2Q^2L^2r_+}, 
%\end{equation}
%or equivalently, via the condition $f(r_+)=0$,
\begin{equation}\label{dr}
\d r_+ = \frac{2r_+^2|p^r|}{r_+^2-Q^2}.
\end{equation}

Now, from the extremal condition $f(r_+)=f'(r_+)=0$, we can express the charge $Q$ of an extremal black hole in terms of the horizon radius $r_+$ and AdS radius $L$ as
\begin{equation}
Q_\text{ext}^2=r_+^2 + \frac{3r_+^4}{L^2}.
\end{equation}
(One can also write the enthalpy of an extremal black hole as
$M_\text{ext} = r_+ + 2r_+^3/L^2$.)
For an extremal black hole then, Eq. (\ref{dr}) can be re-written as
\begin{equation}
\d r_+ = -\frac{2L^2|p^r|}{3r_+^2},
\end{equation}
which is negative. By continuity this means that $\d r_+$ could be negative for \emph{near-extremal} black holes.

Since the entropy of the black hole is $S=A/4=\pi r_+^2$, where $A$ denotes the horizon area, we have $\d S = 2\pi r_+ \d r_+$. It then follows that the second law could be violated. This is the argument in \cite{1709.08665}, which was then repeated in many subsequent follow-up works  \cite{1901.04247, 1901.05589, 1901.08915, 1901.10660, 1902.06489, 1903.03764, 1904.12365,1905.01618,1905.07747,1905.07750}. Our point of view is that this violation of the second law is \emph{not} physical, and should be taken as a \emph{reductio ad absurdum} of the assumption that the infalling particle changes the internal energy by $E=\d U$. Instead we propose that the correct response of the black hole is that its \emph{enthalpy} should change by $E = \d M$. We will explain why this makes physical sense in Sec. (\ref{3}). For now let us demonstrate that assuming $E = \d M$ would ensure the second law is always correct.

Indeed, Eq. (\ref{gwak}) would now become simply
\begin{equation}
\d M = \frac{Q}{r_+}\d Q + |p^r|.
\end{equation}
Repeating the same calculation as above, \emph{mutatis mutandis}, we would obtain the change in the horizon radius
%\begin{equation}
%\d r_+ = \frac{r_+^3 L^2 \left(\frac{|p^r|}{r_+}+\frac{r_+^2}{L^3}\d L\right)}{2r_+^3-ML^2+r_+L^2}.
%\end{equation}
\begin{equation}
\d r_+ = \frac{2L^2 r_+^2 |p^r| + 2r_+^5 \d L/L}{3r_+^4 + L^2 r_+^2 -L^2Q^2}.
\end{equation}
The denominator is exactly equal to $L^2 r_+^3 f'(r_+)$. It is therefore always positive, though approaching zero in the extremal limit.

Note that in the expression for $\d r_+$, the term involving $\d L$ is no longer absent. However, for fixed cosmological constant ($\d L=0$), we see that $\d r_+ >0$ and thus $\d S>0$ for any Reissner-Nordstr\"om-AdS black hole, always respecting the second law.
(If $\d L < - (L^3/r_+^3)|p^r|$, one would need a further increase in $\d M$ for the second law to be obeyed, but that is beyond the scope of this paper.)

\section{Discussion: Why Enthalpy is the Correct Physical Variable}\label{3}

In the previous section, we have shown that if one takes $E=\d M$, instead of $E =\d U$, then there is no violation of the second law in the extended black hole thermodynamics. 

Now let us justify \emph{why} it should be $E =\d M$. For ordinary thermodynamics of a system at constant pressure, it would indeed be the enthalpy rather than the internal energy that is increased by the amount of heat input into the system, and for the black hole, the energy of the particle seems to be the analogue of the heat. Indeed, $E$ goes into changing $M$, by changing both the internal energy $U$, and by ``creating more volume'' while resisting the pressure. Alternatively, one could interpret this as follows: it is because of the $PV$ term in the formula for the internal energy, $U = M - PV$, that when the particle energy $E$ gives an increase in $M$, this increases $V$, so the $P\d V$ term in $\d U = \d M - P\d V$ at constant $P$ decreases the internal energy.  Thus the increase in the internal energy is less than the increase in the enthalpy, $\d M = E$, produced by the particle absorption. Allowing $P$ to vary changes this picture, but by fixing $P$ we can more readily appreciate why $E = \d M$ makes sense. 

There are many notions of  ``energies'' in thermodynamics, in particular the enthalpy and the internal energy are both ``energy'', but it is important to distinguish which is the correct one that changes under particle absorption. A violation of the second law can arise from working with the wrong mass or energy, for example, for the Kerr-(Newman)-AdS black hole, one has to be careful to use the physical conserved mass, not a mere mass parameter \cite{1506.01248}. In the current context, the issue is similar but subtle. The enthalpy really is \emph{the} conserved mass in the black hole spacetime. In the extended black hole thermodynamics we write it as a sum of the internal energy, $U$, and what amounts to a contribution from the cosmological constant, $PV$.  This does not change the fact that the conserved energy of the particle $E$ should add to the conserved mass of the black hole $M$, since the sum is the conserved ADM mass (see Sec. (IV) of the review article \cite{1207.0887}).

The second law of thermodynamics plays a very fundamental role in physics, so much so that Sir Arthur Eddington once wrote \cite{eddington}: \emph{``The law that entropy always increases holds, I think, the supreme position among the laws of Nature. If someone points out to you that your pet theory of the universe is in disagreement with Maxwell's equations -- then so much the worse for Maxwell's equations. If it is found to be contradicted by observation  -- well, these experimentalists do bungle things sometimes. But if your theory is found to be against the Second Law of Thermodynamics I can give you no hope; there is nothing for it but to collapse in deepest humiliation.''}

If the extended black hole thermodynamics indeed suffered from violations of the second law, it would be a warning sign that something is drastically wrong with promoting the cosmological constant to be a variable. Given that a huge literature exists which shows that extended black hole thermodynamics does give rise to a lot of otherwise sensible results \cite{1608.06147}, it is satisfying to find that the apparent violation of the second law is not a sign of the underlying inconsistency of the framework, but rather due to a misinterpretation of the right physical quantity in the thought experiment.

\begin{acknowledgments}
YCO thanks the National Natural Science Foundation of China (grant No.11705162) and the Natural Science Foundation of Jiangsu Province (No.BK20170479) for funding support. 
DNP gratefully acknowledges the support of the Natural Sciences and Engineering Research Council of Canada.
The authors thank 
De-Chang Dai,
Xiao-Mei Kuang,
Robert Mann, 
Donald Marolf, 
Brett McInnes,
James Nester,
and
Rui-Hong Yue
for related and useful discussions. 
\end{acknowledgments}

\end{document}